# First-principles study of the inversion thermodynamics and electronic structure of Fe$M_2X_4$ (thio)spinels ($M$ = Cr, Mn, Co, Ni; $X$ = O, S)

**David Santos-Carballal,[a,*] Alberto Roldan,[a,b] Ricardo Grau-Crespo,[c] Nora H. de Leeuw[a,b,†]**

[a] *Department of Chemistry, University College London, 20 Gordon Street, WC1H 0AJ, London, UK*

[b] *School of Chemistry, Cardiff University, Main Building, Park Place, Cardiff CF10 3AT, UK*

[c] *Department of Chemistry, University of Reading, Whiteknights, RG6 6AD, Reading, UK*

[*] *E-mail address: david.carballal.10@ucl.ac.uk*

[†] *Tel.: +44 (0)29 20870658, e-mail address: DeLeeuwN@cardiff.ac.uk*

Fe$M_2X_4$ spinels, where $M$ is a transition metal and $X$ is oxygen or sulfur, are candidate materials for spin filters, one of the key devices in spintronics. We present here a computational study of the inversion thermodynamics and the electronic structure of these (thio)spinels for $M$ = Cr, Mn, Co, Ni, using calculations based on the density functional theory with on-site Hubbard corrections (DFT+$U$). The analysis of the configurational free energies shows that different behaviour is expected for the equilibrium cation distributions in these structures: $FeCr_2X_4$ and $FeMn_2S_4$ are fully normal, $FeNi_2X_4$ and $FeCo_2S_4$ are intermediate, and $FeCo_2O_4$ and $FeMn_2O_4$ are fully inverted. We have analyzed the role played by the size of the ions and by the crystal field stabilization effects in determining the equilibrium inversion degree. We also discuss how the electronic and magnetic structure of these spinels is modified by the degree of inversion, assuming that this could be varied from the equilibrium value. We have obtained electronic densities of states for the completely normal and completely inverse cation distribution of each compound. $FeCr_2X_4$, $FeMn_2X_4$, $FeCo_2O_4$ and $FeNi_2O_4$ are half-metals in the ferrimagnetic state when Fe is in tetrahedral positions. When *M* is filling the tetrahedral positions, the Cr-containing compounds and $FeMn_2O_4$ are half-metallic systems, while the Co and Ni spinels are insulators. The Co and Ni sulfide counterparts are metallic for any inversion degree together with the inverse $FeMn_2S_4$. Our calculations suggest that the spin filtering properties of the $FeM_2X_4$ (thio)spinels could be modified via the control of the cation distribution through variations in the synthesis conditions.

## I. INTRODUCTION

The electronics industry has been revolutionized over the last four decades due to the continuous miniaturization of integrated circuits. Spintronics, short for spin electronics, has emerged as the basis for the next generation of electronic devices.[1] The concept of spintronics is to take advantage of both the electron charge and spin in solid-state systems, and therefore its applications require magnetic materials with highly spin-polarized electrons at the Fermi energy.[2] This can be achieved by half-metallic ferrimagnets (HMF)[3] with Curie temperature higher than room temperature. The spin-polarized density of states (DOS) of these compounds has a marked asymmetry around the Fermi energy, where one of the spin channels is a conductor while the other one behaves as an insulator,[4] making them electronic spin filters. Spintronic applications are based on spin valves,[5,6] where two HMF layers are sandwiching a non-magnetic layer. In spintronic applications of high efficiency, the resistivity of the spin valve is required to be extremely sensible to the magnetic field (magnetoresistance).[1]

The magnetoresistive behavior,[7,8] and the half-metallic and ferrimagnetic[9–12] nature of the inverse spinel magnetite ($Fe_3O_4$), together with the ubiquity of this iron oxide,[13] indicates its suitability for spintronic applications.[1,14,15] The origin of these properties in $Fe_3O_4$ has been traditionally rationalized in terms of its inverse spinel structure. The ferrimagnetism in $Fe_3O_4$ arises from the antiparallel alignment of the magnetic moments of the ions in the tetrahedral and octahedral sublattices (which is known as collinear Néel configuration),[16] while the hopping of the extra electron in the minority channel of the spins explains the half-metallic properties.[9] Greigite ($Fe_3S_4$) has been shown to have a similar electronic structure to its oxide counterpart $Fe_3O_4$.[10] Both compounds are sometimes found associated with other transition metals of similar ionic radii[17] and valence as Fe, such as Mn, Co, Ni and Cr,[18] forming spinel compounds of formula $FeM_2X_4$.[19] In these systems, $M$ represents the transition metal and $X$ represents the oxygen or sulfur atom, where the sulfide spinels are usually called thiospinels.[20–22] The substituted spinels could retain the type of magnetic behavior of their parent compounds (magnetite or greigite), which is driven by a negative superexchange interaction that is stronger between ions occupying different sublattices than between ions within the same sublattice.[16]

The crystal structure of a (thio)spinel is face-centered-cubic and the space group is $Fd\overline{3}m$. The cubic unit cell contains eight units of $FeM_2X_4$ where the 32 anions are in a cubic closed packed arrangement, while 8 of the tetrahedral sites and 16 of the octahedral ones are occupied by all the cations, see Fig. 1. As originally suggested by Barth and Posnjak,[23] different cation arrangements of the (thio)spinel formula can be rewritten as $\left(\mathrm{Fe}_{1-x}M_x\right)_{\mathrm{A}}\left(\mathrm{Fe}_x M_{2-x}\right)_{\mathrm{B}} X_4$, where A and B denote tetrahedral and octahedral sites respectively, while $x$ is the degree of inversion. In normal spinels ($x = 0$), Fe ions occupy exclusively the A sublattice and $M$ is confined to the B sublattice. In inverse spinels ($x = 1$), the A sublattice holds half of the $M$ cations and the B sublattice is equally populated with Fe and $M$ ions. When $0 < x < 1$, Fe and $M$ have an intermediate degree of distribution within the A and B sublattices. For all inversion degrees, the $Fd\overline{3}m$ symmetry of the spinel is usually retained, as long as all cations are distributed completely randomly within each sublattice, which makes all the sites within each sublattice effectively equivalent. In such cases the cation distribution is fully characterized by the inversion degree $x$. The degree of inversion in spinels has been found to be affected by different factors, including the ionic radii of the distributed species, the electronic configuration, the electrostatic energy of the lattice, the short-range Born repulsion energy, crystal field effects, and polarization effects.[18,19]

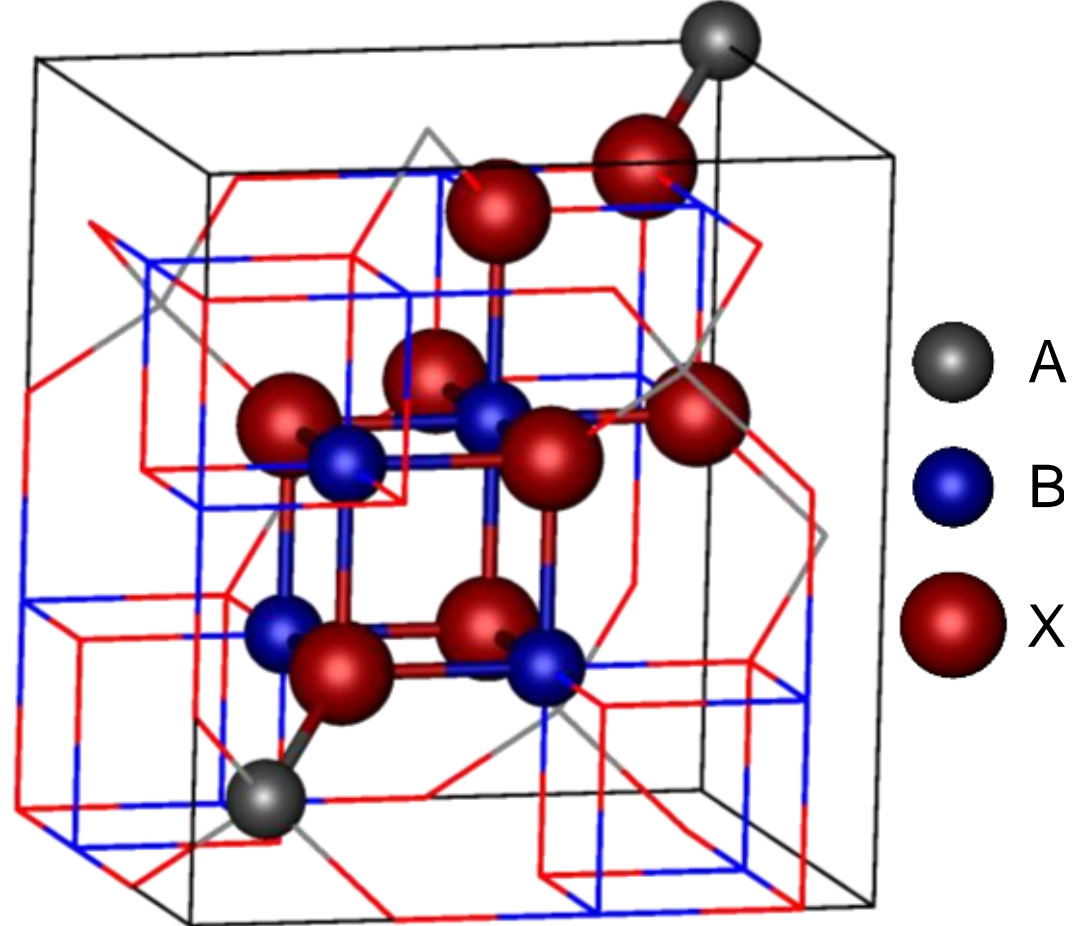

FIG. 1. Schematic representation of one full unit cell of a perfect spinel, highlighting one of the four primitive rhombohedral cells. The spinel structure has the symmetry group $Fd\bar{3}m$ with three ion sites: tetrahedral (A), octahedral (B) cation positions and the anion (X) position.

Structural aspects of $FeM_2X_4$ (thio)spinels have been reported extensively in the literature, sometimes also addressing their influence on the magnetic and electronic properties. For example, these studies include: (1) the mixing, non-stoichiometry[24] and magnetic properties as a function of the cation site distribution of the $Fe_3O_4$-$FeCr_2O_4$ system;[25] (2) the magnetic ordering of $FeCr_2O_4$[26–29] and $FeMn_2O_4$[30] upon crystal symmetry lowering; (3) the relevance of the electronic structure to the magnetic properties of $FeCr_2X_4$;[31] (4) the transport properties based on the half-metallic electronic structure of $FeCr_2S_4$;[32] (5) the magnetic structures in $FeCr_2X_4$[33] and the colossal magnetoresistance in $FeCr_2S_4$;[34] (6) the degree of inversion in $FeMn_2O_4$[35–38] and $FeCo_2O_4$;[39–41] (7) the structural phase stability and magnetism in $FeCo_2O_4$;[42] (8) the structural and magnetic properties of $FeNi_2O_4$,[43] as well as (9) the thermodynamic stability[44] and cation distribution of $FeNi_2S_4$.[45,46] Nevertheless, in compounds such as these (thio)spinels, where the other transition metal's atomic number differs only by 1 from Fe, the X-ray diffraction intensities are very similar for any inversion degree, which makes it difficult experimentally to differentiate the location of the cations in the structure.

Owing to the experimental limitations for the determination of the cation arrangement in $FeM_2X_4$ (thio)spinels, in the present work we have used DFT+$U$ calculations to investigate systematically how modifying the spinel composition affects the equilibrium inversion degree and how this determines the magnetic and electronic properties at a given composition. We study the influence of the nature of the $M$ and $X$ ions ($M$ = Cr, Mn, Co, Ni and $X$ = O, S) on these properties, a type of investigation that has been undertaken previously for other groups of oxide spinels[14] and Heusler alloys[47,48] with potential application in spintronic devices. We will discuss from a thermodynamic point of view the equilibrium cation distribution of these (thio)spinels and the role of the ions' sizes and crystal field stabilization effects. We will also analyze the dependence of the electronic and magnetic structure on the degree of inversion for the normal and completely inverse systems.

## II. COMPUTATIONAL METHODS

### A. Calculation details

We have carried out spin-polarized quantum mechanical calculations using density functional theory (DFT) as implemented in the Vienna *Ab-initio* Simulation Package (VASP).[49–52] The Perdew-Burke-Ernzerhof functional revised for solids (PBEsol)[53] was the version of the generalized gradient approximation (GGA) used as exchange-correlation functional for all geometry optimizations and for the calculation of all density of states (DOS), because PBEsol provides a better description of the structure of solids than its parent functional.[54]

The semiempirical method of Grimme (D2) was also included in our calculations for modelling the long-range van der Waals interactions.[55] Even when these interactions are not expected to affect significantly the bulk properties of the hard solids investigated here, we have included the D2 correction at this stage because in future work we expect to study the surfaces of these solids and their interactions with adsorbates, where dispersion effects may play a significant role.[56–61] The projector augmented wave (PAW) pseudopotential method[62,63] was used to describe the core electrons and their interaction with the valence electrons, *i.e.* those in level 4*d* for Fe, Co and Ni, 3*p*4*d* for Cr and Mn, 2*s*2*p* for O and 3*s*3*p* for S. The kinetic energy cutoff for the plane-wave basis set expansion was set at 520 eV for the geometry optimizations in order to avoid the Pulay stress arising from the cell shape relaxations. A Monkhorst-Pack grid of 7×7×7 $\Gamma$-centred *k*-points[64–66] was used for all calculations. During relaxation, Feynman forces on each atom were minimized until they were less than 0.01 $\text{eV}\cdot\text{Å}^{-1}$. For the calculation of the DOS we applied the tetrahedral method with Blöchl corrections. Atomic charges and atomic spin moments were analyzed using the Bader partition methodology[67] in the implementation of Henkelman *et al.*[68–70]

In order to improve the description of the highly correlated 3*d* electrons in the spinels under study, we have included the Dudarev *et al.*[71] approach for the *d* orbital correction within the DFT + *U* method.[72] We report in Table I the values used for the on-site Coulomb interaction term of *d* Fe and *d M*. These values were determined by fitting the calculated positions of the *d* band centers to those obtained from calculations using the screened hybrid functional of Heyd-Scuseria-Ernzerhof (HSE06),[73–79] which provides band gaps of better quality than semi-local functionals.[80] The HSE06 is made up from the Perdew-Burke-Ernzerhof functional (PBE)[81,82] exchange and correlation components, mixed with 25% of short-range Hartree-Fock (HF) exchange.[73] The Coulomb potential exchange is replaced by a screened potential (with screening parameter $\omega$ = 0.207 Å$^{-1}$) in order to define the separation between the short- and long-range components of the HF exchange.[79] While the amount of short-range HF is a constant determined by perturbation theory, making HSE06 an adiabatic connection functional in this part of the potential,[83] its screening parameter is a reasonable system-averaged value across a wide variety of systems, giving better agreement with experiments in the case of semiconductors than for metals or insulators.[80]

For the fitting, we carried out single-point calculations with both PBEsol + $U$ and HSE06, using unrelaxed structures with normal cation distributions ($a_0$ and $u_0$ were taken from experiment for these calculations, values are listed in Table II). In a first step, we determine $U_{eff}$ for Fe, by considering the $Fe_3O_4$ and $Fe_3S_4$ electronic structures (which have been studied before).[10,44,84] We then keep these values for Fe and perform a set of DFT + $U$ calculations where the effective Hubbard parameter ($U_{eff}$) of the $M$ ion was changed in steps of 0.5 eV from 0 to 6.0 eV. In all the HSE06 calculations, we used the same settings as for the PBEsol simulations.

TABLE I. Summary of the optimum effective Hubbard parameter ($U_{eff}$) in eV used through this work for the spinel oxides and sulfides.

| | Cr | Mn | Fe | Co | Ni |
|---|---|---|---|---|---|
| Fe$M_2$O$_4$ | 4.0 | 3.5 | 4.0 | 1.5 | 5.5 |
| Fe$M_2$S$_4$ | 2.0 | 2.5 | 3.5 | 0.5 | 4.5 |

We found that the optimum $U_{eff}$ values for the Cr, Mn and Fe ions in the spinel oxides are within 0.5 eV of the ones previously found for PBE + $U$ by Wang *et al.*[85] via comparison of experimental and theoretical formation energies of metal oxides. The two exceptions are the $U_{eff}$ values for Co and Ni which differ, according to our methodology, by 1.8 and 0.9 eV respectively from the ones reported by Wang *et al.*[85] The smaller $U_{eff}$ values of the thiospinels, compared to their oxide counterparts, reflect their more covalent character. The $U_{eff}$ for Mn-based thiospinel compares well with the value reported by Rohrbach *et al.*,[86] while, according to our methodology, the one for Fe is 1.5 eV above the one used by the same authors. Hence, in general our $U_{eff}$ are similar to previously employed values, with some differences which can be expected from the use of a different starting functional (PBEsol in our case), implementation of the method, different compound or different fitting procedure.

All the calculations were performed in the rhombohedral primitive unit cell of the $\mathrm{Fe}M_2X_4$ spinels, which comprises 14 atoms, see Fig. 1. For each composition of $\left(\mathrm{Fe}_{1-x}M_x\right)_\mathrm{A}\left(\mathrm{Fe}_xM_{2-x}\right)_\mathrm{B}X_4$, we considered three values of $x$ (0, 0.5 and 1.0). When using this cell, the site occupancy artificially lowers the symmetry from space group $Fd\overline{3}m$ (No. 227) in the normal spinel to *R3m* (No. 160) in the half-inverted and to *Imma* (No. 74) in the fully-inverted spinel.[87] The use of the primitive cell ensures that there is a single cation configuration for each of these three degrees of inversion, which simplifies the simulations, allowing us to scan a wide range of $\mathrm{Fe}M_2X_4$ compositions. This approximation follows previous work, where the use of the primitive cell model has been found to adequately describe experimental properties of half- and fully-inverted spinels.[87–91] However, we cannot rule out completely that the use of larger supercells could actually lead to cation configurations with lower energies for the same inversion degree, as found for example in a recent study of $\mathrm{CoFe_2O_4}$.[92]

Following the collinear Néel model, the initial magnetic moments of the atoms within each sublattice were set parallel among themselves and antiparallel to those of the other sublattice. For each inversion degree, we ran several calculations, specifying different initial magnetic moments, corresponding to different combinations of low- and high-spin states for the transition metal ions in each sublattice, in order to identify the ground state. The magnetic moments were allowed to relax during each of the calculations. It should be noted that the magnetic structure with antiparallel sublattices is not strictly valid in the case of $\mathrm{FeCr_2O_4}$, which is known to have a spiral magnetic structure.[33] However, for the sake of comparison with the other spinel systems, we have not considered its different magnetic structure in this study.

The experimental lattice ($a_0$) and anion ($u_0$) parameters (defining the anion position in the crystal) of the spinels are shown on Table II. These were used as the starting structures for our simulations, where $a_0$ and the internal coordinates were allowed to relax fully for each inversion degree. We kept the cell shape perfectly rhombohedral in such a way that the conventional cell was always cubic. As $FeMn_2S_4$ and $FeCo_2S_4$ spinels have not been characterized so far, we postulated an initial hypothetical structure for both. For the Mn- and Co-based thiospinels, we kept the same initial anion parameter as in their oxide counterparts and scaled up their initial lattice parameter according to the equation:

$$a_0\left(\mathrm{Fe}M_2\mathrm{S}_4\right)=\frac{a_0\left(\mathrm{Fe_3S_4}\right)\cdot a_0\left(\mathrm{Fe}M_2\mathrm{O}_4\right)}{a_0\left(\mathrm{Fe_3O_4}\right)}, \tag{1}$$

which gives the estimates shown in Table II.

TABLE II. Summary of the initial unit cell lattice ($a_0$) and anion ($u_0$) parameters of $FeM_2X_4$ spinels used in this work. The relaxed $a$ and $u$ are also reported for $x$ = 0, 0.5 and 1. Note that the origin is the center of symmetry.[19]

| Structure | Experimental | | | $x$ = 0 | | $x$ = 0.5 | | $x$ = 1 | |
|---|---|---|---|---|---|---|---|---|---|
| | $a_0$ (Å) | $u_0$ | Reference | $a$ (Å) | $u$ | $a$ (Å) | $u$ | $a$ (Å) | $u$ |
| $FeCr_2O_4$ | 8.38 | 0.261 | 33 | 8.351 | 0.261 | 8.372 | 0.261 | 8.392 | 0.265 |
| $FeCr_2S_4$ | 10.00 | 0.259 | 33 | 9.830 | 0.258 | 9.855 | 0.262 | 9.898 | 0.260 |
| $FeMn_2O_4$ | 8.51 | 0.250 | 93 | 8.420 | 0.256 | 8.436 | 0.265 | 8.446 | 0.267 |
| $FeMn_2S_4$ | 10.04 | 0.250 | This paper* | 9.911 | 0.255 | 9.949 | 0.260 | 9.983 | 0.263 |
| $Fe_3O_4$ | 8.390 | 0.255 | 94 | -- | -- | -- | -- | -- | -- |
| $Fe_3S_4$ | 9.88 | 0.251 | 95 | -- | -- | -- | -- | -- | -- |
| $FeCo_2O_4$ | 8.24 | 0.259 | 40 | 8.196 | 0.256 | 8.168 | 0.259 | 8.119 | 0.271 |
| $FeCo_2S_4$ | 9.73 | 0.259 | This paper* | 9.363 | 0.263 | 9.286 | 0.260 | 9.297 | 0.250 |
| $FeNi_2O_4$ | 8.29 | 0.258 | 43 | 8.124 | 0.259 | 8.123 | 0.260 | 8.133 | 0.255 |
| $FeNi_2S_4$ | 9.47 | 0.257 | 45 | 9.463 | 0.260 | 9.438 | 0.257 | 9.396 | 0.251 |

* Experimental information is not available. Parameters are obtained by scaling the values for the corresponding oxides.

## B. Configurational free energy of inversion

The calculation of the inversion degree in spinels containing two different cations is based on the thermodynamic considerations of Navrotsky and Keppla,[96] which have proved to agree well with experiments.[88,97–100] This methodology is based on the treatment of the spinels' cation distribution as a chemical equilibrium. We calculated the configurational free energy of inversion per formula unit $\Delta F_{\mathrm{config}}$ as,

$$\Delta F_{\mathrm{config}}=\Delta E_{\mathrm{config}}-T\cdot\Delta S_{\mathrm{config}}, \tag{2}$$

where $\Delta E_{\text{config}}$ is the inversion energy per formula unit, $T$ is the temperature and $\Delta S_{\text{config}}$ is the ideal configurational entropy also per formula unit, which is calculated as,

$$\Delta S_{\text{config}} = -R\left[x\ln x + (1-x)\ln(1-x) + x\ln \tfrac{x}{2} + (2-x)\ln\left(1-\tfrac{x}{2}\right)\right], \quad (3)$$

where $R$ is the ideal gas constant. $\Delta S_{\text{config}}$ = 0 and 11.59 J·mol$^{-1}$·K$^{-1}$ for $x$ = 0 and 1 respectively, while it reaches the maximum value 15.88 J·mol$^{-1}$·K$^{-1}$ for the complete random distribution at $x$ = 2/3. The above expression means that we have only considered ideal contributions to the configurational entropy, in line with previous work.[88,97,100] We are also ignoring vibrational contributions to $\Delta F$, as their contributions are typically small compared to configurational energies and entropies.[88,100]

## III. RESULTS AND DISCUSSION

### A. Equilibrium structures

Table II shows the optimized $a$ and $u$ for the three inversion degrees considered ($x$ = 0, 0.5 and 1.0). In general, the optimized lattice parameter is within 2% from the experimental value, where this is available. However, the relaxed lattice parameter for $FeCo_2S_4$, in the best case ($x$ = 0), is 3.8% away from the initial estimated value, which may be an artefact due to the assumption of linearity between the lattice constants of $Fe_3O_4$, $FeCo_2O_4$ and their sulfide counterparts. After relaxation of the structures, $u$ was still different from the value of ¼ that it has in the perfect spinel. This deviation reflects the displacement, in the (111) direction, of the anion from its ideal position in order to accommodate cations of specific volume. The biggest deviation in $u$ in comparison with the experimental value was for the inverse Mn- and Co-based oxide spinels, Table II. We see that in general, $u$ and $a$ values are sensitive to the cation distribution, although no systematic rule can be derived from their dependence.

### B. Equilibrium inversion degrees

The configurational inversion energy per formula unit ($\Delta E_{\text{config}}$) was fitted *versus* the degree of inversion ($x$) using a quadratic regression curve, see Fig. 2 (a). More details regarding the empirical relationship[101] and theoretical justification[102] of the above fitting in terms of $x$, $a$ and $u$ can be found elsewhere. In this fitting, we defined the normal spinel as the standard state for a given condition of temperature and pressure.

Using the quadratic equation for $\Delta E_{\text{config}}$, it is possible to interpolate the inversion energy for any value of $x$ between 0 and 1. Based on this protocol, we have also estimated the configurational free energy of inversion for a typical firing temperature of 1000 K[31,42,43,45] (among the known cases in this study, $FeNi_2S_4$ is an exception, as it is usually prepared at 573 K[45,103] because it decomposes at 734 K[104]) by using equations (2) and (3). Compounds are usually quenched after synthesis at the firing temperature, retaining the equilibrium cation distribution. We analysed the $\Delta E_{\text{config}}$ dependence with $x$ and provided the equilibrium values of $x$, *i.e.* the ones that satisfy $\partial F_{\text{config}} / \partial x = 0$ at 1000 K, see Fig. 2 (b).

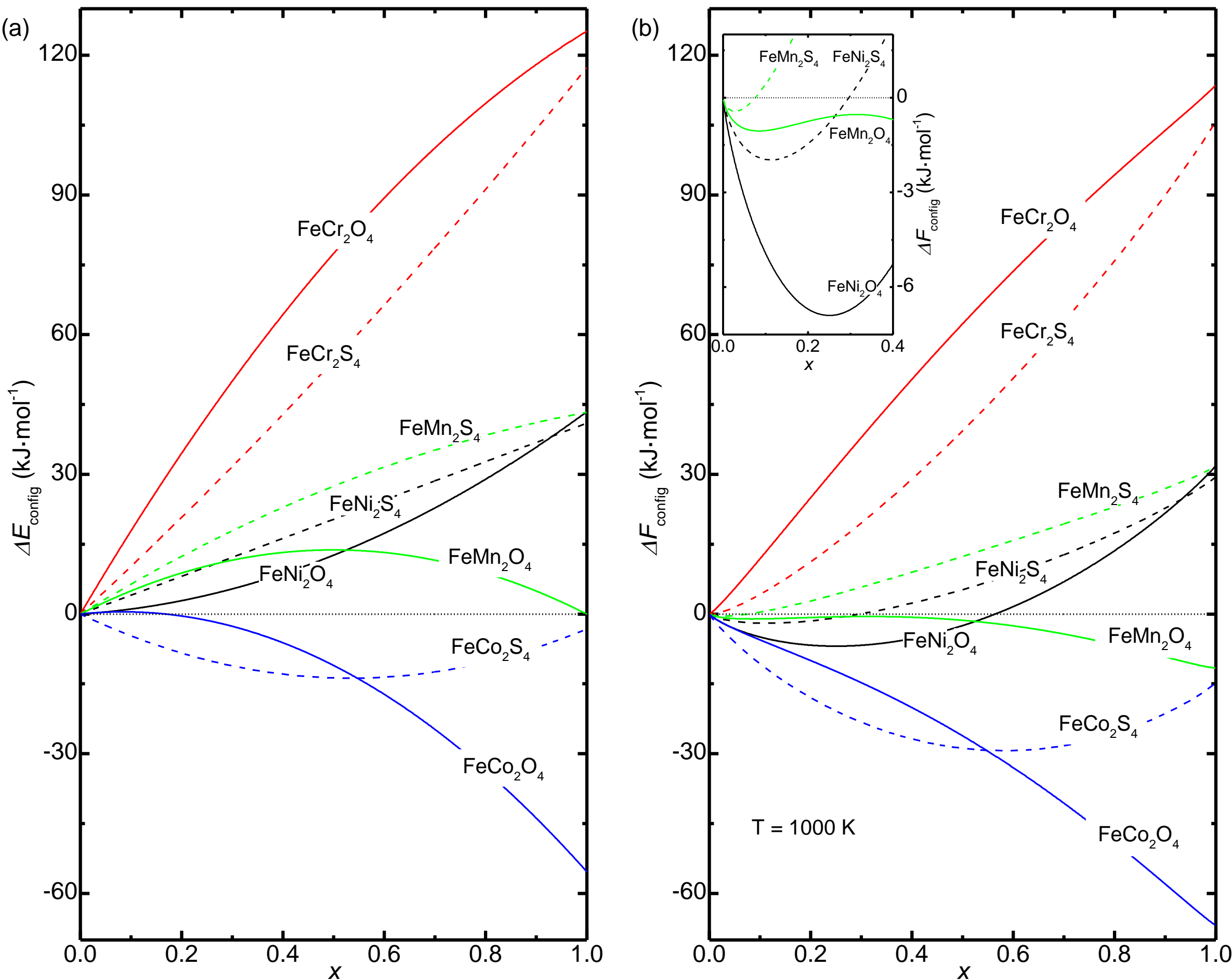

FIG. 2. (a) Configurational inversion energy ($\Delta E_{\text{config}}$) and (b) configurational inversion free energy ($\Delta F_{\text{config}}$) as a function of the inversion degree for Fe$M_2X_4$ spinels. Inset shows enlargement of $\Delta F_{\text{config}}$ for the FeMn$_2X_4$ and FeNi$_2X_4$ (thio)spinels.

We found the minimum of $\Delta E_{\text{config}}$ to correspond to a normal distribution of cations, with the exception of Co-based systems and $FeMn_2O_4$, Fig. 2 (a). The lowest value of $\Delta E_{\text{config}}$ for $FeCo_2O_4$ spinel is found to be an inverse cation distribution, whereas for $FeMn_2O_4$, both normal and inverse cation distribution structures lie at similar energies, while the intermediate degree of inversion ($x = 0.5$) is only ~14 $\text{kJ} \cdot \text{mol}^{-1}$ above the ground state. $FeCo_2S_4$ is an atypical thiospinel in this study, in the sense that it shows a critical point of low energy with intermediate cation distribution, at $x = 0.53$.

**FeCr$_2$*X*$_4$.** From Fig. 2 (b), we deduced that at 1000 K, the Cr-based (thio)spinels are normal under equilibrium conditions, as a result of a highly endothermic process of inversion. This normal cation distribution of $FeCr_2X_4$ is supported by powder neutron diffraction intensities[33] at room temperature and for the oxide by Mössbauer measurements[25] and DFT calculations,[24] see Table III.

TABLE III. Summary of equilibrium inversion degree ($x$) of $FeM_2X_4$ (thio)spinels from this work and previous reports.

| | $x$ | Reference | $x$ at 1000 K |
|---|---|---|---|
| $FeCr_2O_4$ | ~ 0.0 | 24,33 | 0.00 |
| | 0.00 | 25 | |
| $FeCr_2S_4$ | ~ 0.0 | 33 | 0.00 |
| $FeMn_2O_4$ | 0.5 | 35,36 | 0.10 and 1.00 |
| | 0.91 | 37,38 | |
| $FeMn_2S_4$* | -- | -- | 0.03 |
| $FeCo_2O_4$ | 0.52 | 105 | 1.00 |
| | 0.54 | 40 | |
| | 0.565 | 39 | |
| | 0.605 | 106 | |
| | 0.7 | 107 | |
| | 1.0 | 108,109 | |
| $FeCo_2S_4$* | -- | -- | 0.48 |
| $FeNi_2O_4$* | -- | -- | 0.25 |
| $FeNi_2S_4$ | ~ 0.05 | 110 | 0.12 |
| | 0.82 or 1.00 | 21 | |
| | 1.00 | 22,45,46 | |

* Experimental information is not available.

**$FeMn_2X_4$.** At 1000 K, the scenario for $FeMn_2O_4$ is unique in this study, as in addition to the global minimum of $\Delta F_{config}$ at $x = 1$, it has a local one at $x = 0.1$. The local minimum is within a portion of shallow inversion free energy ($0 < x < 0.3$), which may lead to a metastable inversion degree anywhere within this range, for this spinel's equilibrium structure, see Fig. 2 (b). The behaviour of this thermodynamic property in $FeMn_2O_4$ can be rationalized in terms of the small change of $\Delta E_{config}$ with $x$ as well as in $E_{x=0} \approx E_{x=1}$. The upper limit ($x = 0.3$) of the shallow inversion free energy that we predicted agrees semiquantitatively with the experimental inversion degree ($x = 0.5$) found for $FeMn_2O_4$ in a conductivity and thermopower[35] investigation, as well as inferred from the study of a series of spinels,[36] see Table III. The inversion degree has also been found in neutron diffraction experiments to be at $x = 0.91$[37,38] which is in reasonably good agreement with the global minimum calculated here. We speculate that the two inversion degrees of $FeMn_2O_4$ may be hampered by kinetic control, a situation that is outside the scope of this paper, but which explains the different cation arrangements described in the literature. According to our calculations, samples of $FeMn_2O_4$ synthesized above 1150 K can only have inverse cation distribution, as the metastable inversion degree vanishes. $FeMn_2S_4$, on the other hand, is predicted to be mostly normal ($x$ = 0.03) under equilibrium conditions, Fig. 2 (b).

**$FeCo_2X_4$.** $FeCo_2O_4$ is the only completely inverse spinel under equilibrium conditions, due to the highly exothermic process of inversion, which agrees with experimental evidence.[108,109] Nevertheless, our results for the Co-based oxide disagree with the equilibrium inversion degree of $x$ = 0.565 and 0.605 obtained by means of fitting the dependence of the magnetic moment with $x$[39,106] and the similar values within the range $0.52 \leq x \leq 0.7$ derived from Mössbauer spectra,[40,105,107] see Table III. Its sulfide counterpart, which has not been studied experimentally, shows an equilibrium inversion degree of $x$ = 0.48 in our calculations.

**$FeNi_2X_4$.** $FeNi_2O_4$ (and to a lesser extent $FeNi_2S_4$) is predicted to have an intermediate distribution of the cations under equilibrium conditions of around $x$ = 0.25 ($x$ = 0.12 for the thiospinel case). Our results agree with suggestions of partially inverted $FeNi_2S_4$, based on a high temperature calorimetry study of natural samples,[110] see Table III. However, they disagree with the more recent description of synthetic $FeNi_2S_4$ samples as completely inverse spinel based on a neutron powder diffraction measurements at temperatures between 100 and 573 K,[45] thermodynamic-based modelling,[46] EXAFS experiment[22] and Mössbauer data.[21] Based on our calculations and the fact that synthetic $FeNi_2S_4$ samples cannot be annealed to temperatures higher than 734 K,[104] as they decompose, we propose a rationalization of the different cation arrangements found in natural and synthetic samples of this mineral. We suggest that synthesis produces a kinetic product (with $x \sim 1.00$) and that these conditions cannot reproduce the hypogene processes occurring in the ores deep below the Earth's surface that lead to the thermodynamic product found in natural samples ($x \sim 0$).

### C. Size of ions and crystal field stabilization effects

We analyze now the effect of cation size and crystal field stabilization energy on the distribution of cations under equilibrium conditions.

Assuming the hard-sphere model, where the ions are spherical, rigid and in contact, the ratio between the tetrahedral ($R_A$) and octahedral ($R_B$) bond distances will depend solely on $u$. The tetrahedral holes are smaller than the octahedral ones for $u < 0.2625$.[111] Taking into account that for most systems under study here, $u$ is below that value (with a few exceptions in the relaxed structures, see Table II), we can consider that $R_A < R_B$ is expected for a stable spinel.

According to the Shannon effective radii,[17] which depend on the coordination number and oxidation state, $Fe_A^{2+}$ cation has bigger radius than $Cr_B^{3+}$, $Co_B^{3+}$ and $Ni_B^{3+}$ leading to an inverse cation distribution. This agrees very well with our thermodynamic DFT + $U$ calculations for $FeCo_2O_4$ and moderately with the partially inverse ($x = 0.48$) $FeCo_2S_4$ spinel. However, we found the opposite equilibrium distribution for the $FeCr_2X_4$ system and a very small inversion degree ($x < 0.25$) for the $FeNi_2X_4$, indicating that this factor is not the key parameter governing the inversion degree in these compounds. On the other hand, the $Mn_B^{3+}$ Shannon radius is bigger than that of $Fe_A^{2+}$, predicting a normal (thio)spinel. Yet, whereas the sulfide compound is a completely normal spinel, the oxide has a local minimum for $0 < x < 0.3$ and the global one at $x = 1$.

Since we are dealing with open shell $d$ transition metals, the crystal field is also an important effect to consider. McClure,[112] and independently Dunitz and Orgel,[113] derived the crystal field stabilization energy for ions (in oxides) with tetrahedral and octahedral environments, to estimate the relative stability of normal and inverse spinels. Subtracting the tetrahedral stabilization energy from the octahedral one (octahedral site preference energy − OSPE) gives an idea of the octahedral site preference. The OSPE for $Fe^{2+}$ (16.3 kJ/mol) is smaller than for the rest of the cations under consideration here, *i.e.* 195.4 kJ/mol for $Cr^{3+}$, 105.9 kJ/mol for $Mn^{3+}$, 79.5 kJ/mol for $Co^{3+}$ and 95.4 kJ/mol for $Ni^{2+}$ (note that to date no estimation of OSPE for $Ni^{3+}$ is reported). These OSPEs clearly show the preference for normal spinels.

The ambiguities in our results are probably not surprising, because earlier attempts to correlate cation distribution with their size and crystal field effects were also not successful, or at least, unable to provide a complete prediction of the degrees of inversion.[114]

### D. Atomic spin moments and charges

In this and the next section we analyze the electronic and magnetic properties of the spinel materials for the extreme cases of $x = 0$ and $x = 1$.

The total magnetization of saturation ($M_S$) is defined experimentally as the maximum magnetic moment per formula unit of a compound under an increasing magnetic field. This magnitude can also be calculated according to the Néel model as the sum of the atomic spin densities ($m_s$) in the tetrahedral and octahedral sublattices per formula unit.[16] Table IV shows the atomic spin densities for all the compounds under study here. When $x = 0$ in the oxide spinels, $m_s(Fe_A)$ is around 4 $\mu_B$/atom, which is in good agreement with a high-spin electronic distribution for $Fe_A^{2+}: e_{\uparrow}^2 e_{\downarrow}^1 t_{2\uparrow}^3$, with the exception of the deviation in the Cr-based compound. For the normal thiospinels there is more consistency in the $m_s(Fe_A)$ values, as they lie in the range −3.41 and −3.53 $\mu_B$/atom and only in semiquantitative agreement with the Néel model. For the *M* cation in the normal Cr- and Mn-based (thio)spinels, the atomic spin densities are also in good agreement with high-spin electronic distributions. We found that our DFT+*U* calculations underestimated the atomic spin moment of $FeCo_2O_4$ (when $x = 0$) by 1.28 $\mu_B$/atom compared with the one expected from the Néel model for the high-spin distribution of $Co_B^{3+}: t_{2g\uparrow}^3 t_{2g\downarrow}^1 e_{g\uparrow}^2$. In its sulfide counterpart with normal distribution, our calculated value compares well with the one predicted from a low-spin distribution of $Co_B^{3+}$, which renders the ions as non-magnetic. In the case of $\mathrm{Ni}_{\mathrm{B}}^{3+}$, we also found it to be low spin $t_{2g\uparrow}^{3}t_{2g\downarrow}^{3}e_{g\uparrow}^{1}$, although our results overestimated by 0.39 $\mu_B$/atom and underestimated by 0.10 $\mu_B$/atom the expected value for the oxide and sulfide respectively. This agrees with the low-spin cation occupying the octahedral positions in violarite, interpreted previously as $Fe^{2+}$.[21]

TABLE IV. Atomic spin density per atom ($m_s$) and total spin magnetization of saturation per formula unit ($M_S$) both calculated by means of a Bader analysis and in $\mu_B$.

| Spinel | $x$ | $m_s$ | | | | $M_S$ | $m_s$ | | | | $M_S$ |
|---|---|---|---|---|---|---|---|---|---|---|---|
| | | A | B1 | B2 | $X$ = O | | A | B1 | B2 | $X$ = S | |
| $FeCr_2X_4$ | 0 | -3.72 | 2.92 | -- | -0.03 | 2.00 | -3.46 | 2.95 | -- | -0.11 | 2.00 |
| $FeCr_2X_4$ | 1 | -3.50 | 2.82 | 4.09 | 0.15 | 4.00 | -3.28 | 2.86 | 3.79 | 0.16 | 4.00 |
| $FeMn_2X_4$ | 0 | -3.97 | 4.16 | -- | -0.09 | 4.00 | -3.53 | 4.03 | -- | -0.13 | 4.00 |
| $FeMn_2X_4$ | 1 | -4.49 | 3.73 | 4.17 | 0.15 | 4.00 | -4.17 | 3.94 | 3.84 | 0.10 | 4.02 |
| $FeCo_2X_4$ | 0 | -3.95 | 2.72 | -- | 0.13 | 2.00 | -3.45 | -0.04 | -- | -0.08 | -3.87 |
| $FeCo_2X_4$ | 1 | -2.44 | 0.01 | 4.11 | 0.08 | 2.00 | -0.91 | 0.03 | 3.32 | 0.00 | 2.46 |
| $FeNi_2X_4$ | 0 | -4.04 | 1.39 | -- | -0.18 | -2.00 | -3.41 | 0.90 | -- | -0.02 | -1.69 |
| $FeNi_2X_4$ | 1 | -1.89 | 1.59 | 4.10 | 0.05 | 4.00 | -0.52 | 0.86 | 3.50 | 0.04 | 4.00 |

In the inverse (thio)spinels, the calculated spin densities for $\mathrm{Fe}_{\mathrm{B2}}^{2+}$ were slightly overestimated in the oxides compared with the Néel model, while they were more moderately underestimated in the sulfides considering a high-spin electronic distribution of $t_{2g\uparrow}^{3}t_{2g\downarrow}^{1}e_{g\uparrow}^{2}$ for these ions, see Table IV. The calculated atomic spin densities of the inverse Cr- and Mn-based (thio)spinels agree better, especially in the cations occupying the octahedral (B1) positions, with the high-spin electronic distribution for these ions, as described for the normal spinels. However, in the Co- and Ni-based inverse compounds, we found low-spin densities for these atoms in the B1 positions, where the nearly diamagnetic $\mathrm{Co}_{\mathrm{B1}}$ in the inverse $FeCo_2O_4$ agrees with experiments.[108,109] Notable exceptions are the $\mathrm{Co}_{\mathrm{A}}^{3+}$ in the thiospinel and $\mathrm{Ni}_{\mathrm{A}}^{3+}$ in both oxide and sulfide compounds, where our calculations shift $m_s$ by more than 1 $\mu_B$/atom with the expected value (in the best case) from a low-spin electronic distribution for these atoms.

The most stable normal cation distribution of $FeCr_2O_4$ gave the closest $M_S$ to the experimental one, although still overestimated by 1.35 $\mu_B$/f.u., as this measurement was carried out at a temperature in which the spins are not collinear anymore.[33] In the case of its normal sulfide counterpart, the difference in spin magnetization of saturation with the experiments is smaller (0.41 $\mu_B$/f.u.).[33]

The Ni-based (thio)spinels are found experimentally to be paramagnetic.[21,43,103] In the oxide this has been explained as being due to high-spin $Fe^{3+}$ ions exclusively localized on the A sublattice whose spins compensate completely the [$Ni^{2+}Ni^{3+}$] occupying the octahedral positions.[43] In the sulfide this has been rationalized on the basis of an A sublattice filled by $Ni^{3+}$ and low-spin $Fe^{2+}$ occupying octahedral positions.[21] Here, based on our calculated spin magnetization of saturation and assuming intermediate degrees of cation distribution, we present a fresh explanation for the paramagnetism of $FeNi_2X_4$. Considering that $M_S$ changes linearly with $x$, we may postulate that the oxide and sulfide will be paramagnetic when $x$ = 0.33 and 0.30 respectively. Although this suggestion agrees with the oxide and sulfide equilibrium inversion degree calculated in section 3.1, it shows that paramagnetism in these compounds may be due not to the canonical inverse spinel structure with integer oxidation numbers, but to intermediate inversion degrees.

To the best of our knowledge, there is no experimental determination of the saturation magnetization of either $FeCo_2S_4$ or $FeMn_2S_4$. Although we found both compounds to be ferrimagnetic, the measurement of $M_S$ for $FeCo_2S_4$ may be essential to determine the inversion degree of this spinel, as our calculation of the normal and inverse cation distributions show different magnetizations of saturation. Our results agree with the ferrimagnetic behaviour described for $FeCo_2O_4$[115] and $FeMn_2O_4$,[30] below the Curie (Néel) temperature.

Table V shows the charges ($q$) gained or lost by an atom with respect to the neutral atom in the $FeM_2X_4$ spinels. We clearly appreciate, that charges are systematically underestimated for all the $FeM_2X_4$ (thio)spinels. For the Cr- and Mn-based (thio)spinels, $q_A$ is clearly smaller than $q_B$ for any inversion degree and also for the inverse $FeCo_2O_4$. However, for the Co- and Ni-based systems, the relative charges of the atoms in the tetrahedral and octahedral positions is different in the oxide and sulfide. In the spinel oxides, together with the normal Co- and Ni- thiospinels, $q_A$ is bigger than $q_B$, while in the inverse thiospinels, it is the other way round.

TABLE V. Calculated Bader charges in the Fe$M_2X_4$ spinels.

| Spinel | $x$ | A | B1 | B2 | $X$ = O | A | B1 | B2 | $X$ = S |
|---|---|---|---|---|---|---|---|---|---|
| $FeCr_2X_4$ | 0 | 1.33 | 1.75 | -- | -1.21 | 0.92 | 1.22 | -- | -0.84 |
| $FeCr_2X_4$ | 1 | 1.47 | 1.76 | 1.58 | -1.20 | 1.11 | 1.22 | 1.12 | -0.86 |
| $FeMn_2X_4$ | 0 | 1.50 | 1.56 | -- | -1.16 | 0.95 | 1.18 | -- | -0.83 |
| $FeMn_2X_4$ | 1 | 1.41 | 1.69 | 1.68 | -1.20 | 1.08 | 1.21 | 1.17 | -0.86 |
| $FeCo_2X_4$ | 0 | 1.52 | 1.34 | -- | -1.05 | 0.86 | 0.54 | -- | -0.49 |
| $FeCo_2X_4$ | 1 | 1.30 | 1.31 | 1.68 | -1.07 | 0.46 | 0.60 | 0.95 | -0.50 |
| $FeNi_2X_4$ | 0 | 1.61 | 1.18 | -- | -0.99 | 0.86 | 0.57 | -- | -0.50 |
| $FeNi_2X_4$ | 1 | 1.27 | 1.16 | 1.67 | -1.03 | 0.26 | 0.61 | 0.99 | -0.46 |

### E. Electronic density of states

#### *1. $FeCr_2X_4$*

The density of states (DOS) in Fig. 3 show that at $x$ = 0, $FeCr_2X_4$ is half-metallic, which we confirmed by the integer value of total spin magnetization ($M_S$ = 2.00 $\mu_B$/f.u.), see Table IV. An integer value of the total spin magnetization discriminates half-metals and insulators from metals. The total number of electrons of any stoichiometric system is integer ($n$) and if it has a band gap at least in one spin channel, there is an integer number of electrons ($n'$) there too. This makes the difference ($n'' = n - n'$), which is the number of electrons on the band that crosses the Fermi level also integer. Therefore, the magnetization of saturation, *i.e.* the difference of $n'$ and $n''$, is also integer.[3,15,116,117]

The DOS shows a sharp peak of the partially-occupied $e$ level of $Fe_A$ ions in the majority spin channel (α) crossing the Fermi energy, which is weakly hybridized with the empty Cr $e_g$ level in the oxide spinel, while the minority spin channel (β) shows a gap near $E_F$. There is a nearly equally intense band due to the occupied $Cr_B$ $t_{2g}$ level in the majority channel of the spins at −3.0 eV in the oxide (−1.75 eV in the sulfide), which suggests that the half-metallic properties do not involve the sublattice B. In the oxide, the other valence bands of the $Fe_A$ ions ($t_2$ and $e$ levels) appear in the minority channel of the spins below −2.5 eV, always strongly hybridized with the O 2*p* orbitals. However, the Cr $t_{2g}$ level, together with a small contribution from the $e_g$ orbitals, in the valence part of the majority spin channel are weakly hybridized with the O 2*p* orbitals. The unoccupied $t_2$ level of $Fe_A$ appears at 3.0 eV in the majority channel of the spins while $Cr_B$ has the unoccupied $t_{2g}$ level in the majority channel of the spins (1.5 eV) and the $t_{2g}$ and $e_g$ levels in the minority channel of the spins (2.5 and 4.0 eV).

The inversion of half of the Cr cations to the tetrahedral positions in $FeCr_2X_4$, generates four non-equivalent types of atoms (B1 and B2 are the two types of atoms occupying B positions), see Fig. 3 right panels. With this cation distribution, the (thio)spinel is still half-metallic ($M_S$ = 4.00 $\mu_B$/f.u., see Table IV), but unlike in the normal spinel structure, through the minority channel of the spins due to the partially-occupied $t_2$ level of the $Cr_A$ ions. All the $Cr_A$ bands are shifted towards higher energy values with respect to $Fe_A$ in the normal (thio)spinel. The $Cr_{B1}$ *d* bands appear roughly in the same position as in the normal (thio)spinel, although less intense. The $e_g$ and $t_{2g}$ levels of $Fe_{B2}$, which lie very close or are hybridized, are around −8.0 and 1.5 eV in the α and β channel of the spins, respectively. As a result of the cations' shifted and split bands, there is less hybridization of the O 2*p* orbitals in the valence regions compared to the normal spinel, which are prominent in this section. The main difference, for any cation distribution, between the DOS of $FeCr_2S_4$ and its oxide counterpart is that all the bands in the sulfide are squeezed towards the Fermi energy.

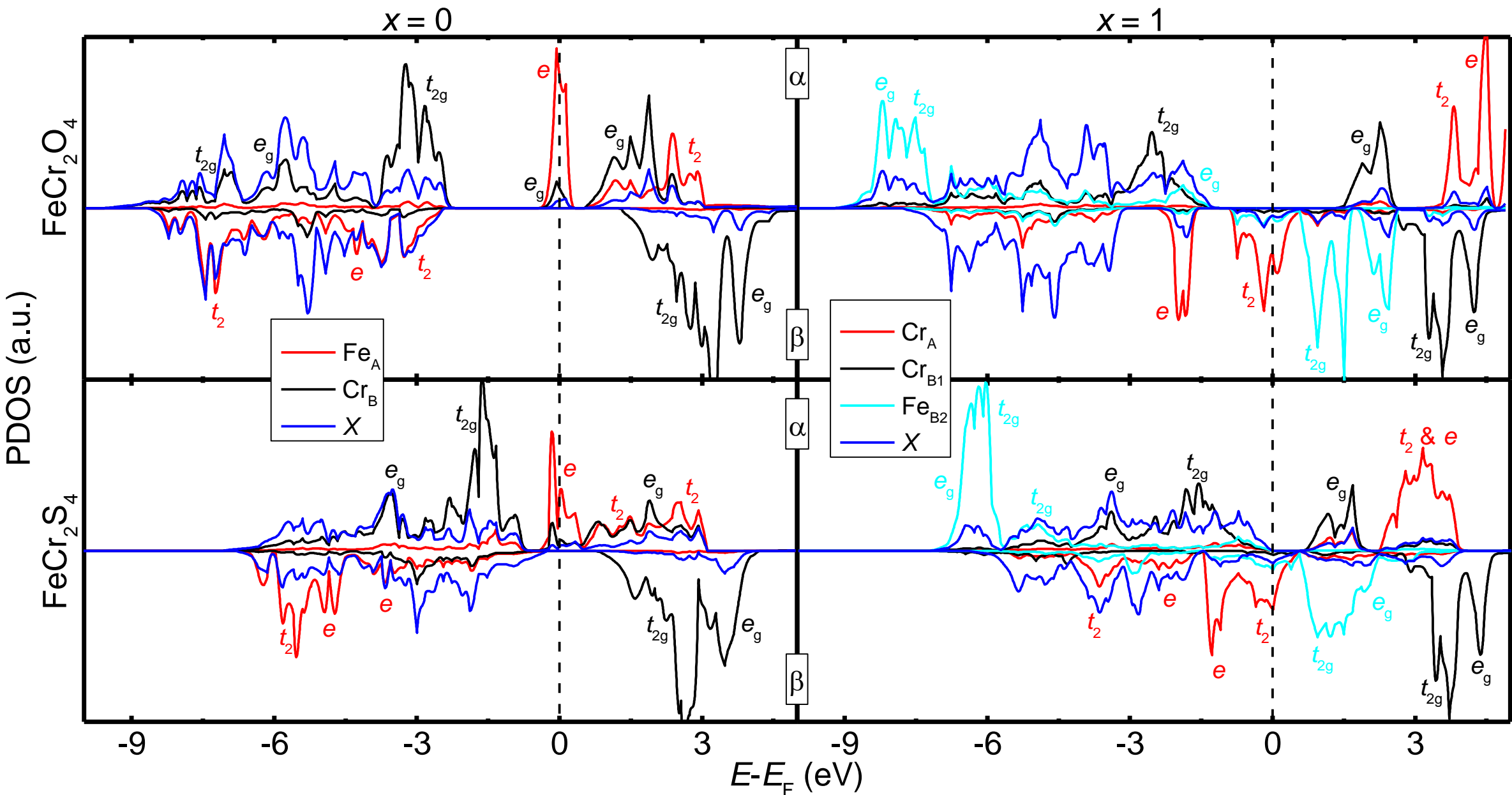

FIG. 3. Atomic projections of the spin decomposed total density of states (PDOS) for $FeCr_2O_4$ and $FeCr_2S_4$. Fe and Cr contributions are from the 3*d* bands. O and S contributions are from the 2*p* and 3*p* orbitals respectively.

### *2. $FeMn_2X_4$*

When the $FeMn_2X_4$ (thio)spinel is normal ($x$ = 0), the $Fe_A$ $e$ and $t_2$ levels, which are very close or hybridized, appear in the maximum of the valence and minimum of the conduction bands in the minority and majority channel respectively, see Fig. 4 left panels. The half-metallic character of the normal $FeMn_2X_4$ (thio)spinels is also confirmed by the spin density analysis, showing a spin magnetization per formula unit of $M_S$ = 4.00 $\mu_B$, see Table IV. At the Fermi energy, the spin up partially-occupied $e_g$ band of $Mn_B$ appears highly hybridized with the $X$ $p$ orbitals and with the $Fe_A$ $e$ and $t_2$ levels in the oxide and sulfide respectively, in agreement with the bigger atomic volume, enhancing orbital overlapping. The rest of the density of states is essentially the same as in the Cr-material, while in the Mn-based spinels the valence band is slightly shifted towards the Fermi energy and the $Fe_A$ $t_2$ and $e$ bands in the β channel of the spins appear more prominently. With the normal cation distribution, the bands of $FeMn_2S_4$ (Fig. 4 bottom-left panel), as in the case of $FeCr_2S_4$, are shifted towards the Fermi energy, except for the $Mn_B$ $t_{2g}$ and $e_g$ bands in the α channel of the spins.

When $FeMn_2O_4$ is a completely inverse spinel ($x$ = 1), the system becomes half-semiconductor with a negligible band-gap, which has also been found experimentally,[35] see Fig. 4 top-right panel and also note in Table IV the integer $M_S$ = 4.00 $\mu_B$/f.u., typical of materials with band gaps. The partially-occupied and split $Mn_{B1}$ $e_g$ bands appear close to the Fermi energy in the majority channel of the spins, where the $t_{2g}$ level, with a small contribution from the $e_g$ level, in the conduction band is highly hybridized with the O $p$ orbitals. The $Mn_A$ $t_2$ and $e$ levels in the minority channel of the spins, are merged altogether and appear as a wide conduction band. In the inverse $FeMn_2S_4$ spinel, the bands are squeezed towards the Fermi energy, becoming a metal in both channels of the spins. The bands responsible for the conductivity properties are associated with the $Mn_{B1}$ $e_g$ level and with the $Fe_{B2}$ $t_{2g}$ level in the majority and minority channel of the spins respectively, see Fig. 4 bottom-right panel and the decimal $M_S$ = 4.02 $\mu_B$/f.u., typical of metals in Table IV. In general, we see that for any inversion degree, upon exchange of Cr by Mn cations, the bands responsible for the conduction properties are no longer the ones belonging to the atoms occupying the tetrahedral positions but those of $Mn_{B(1)}$.

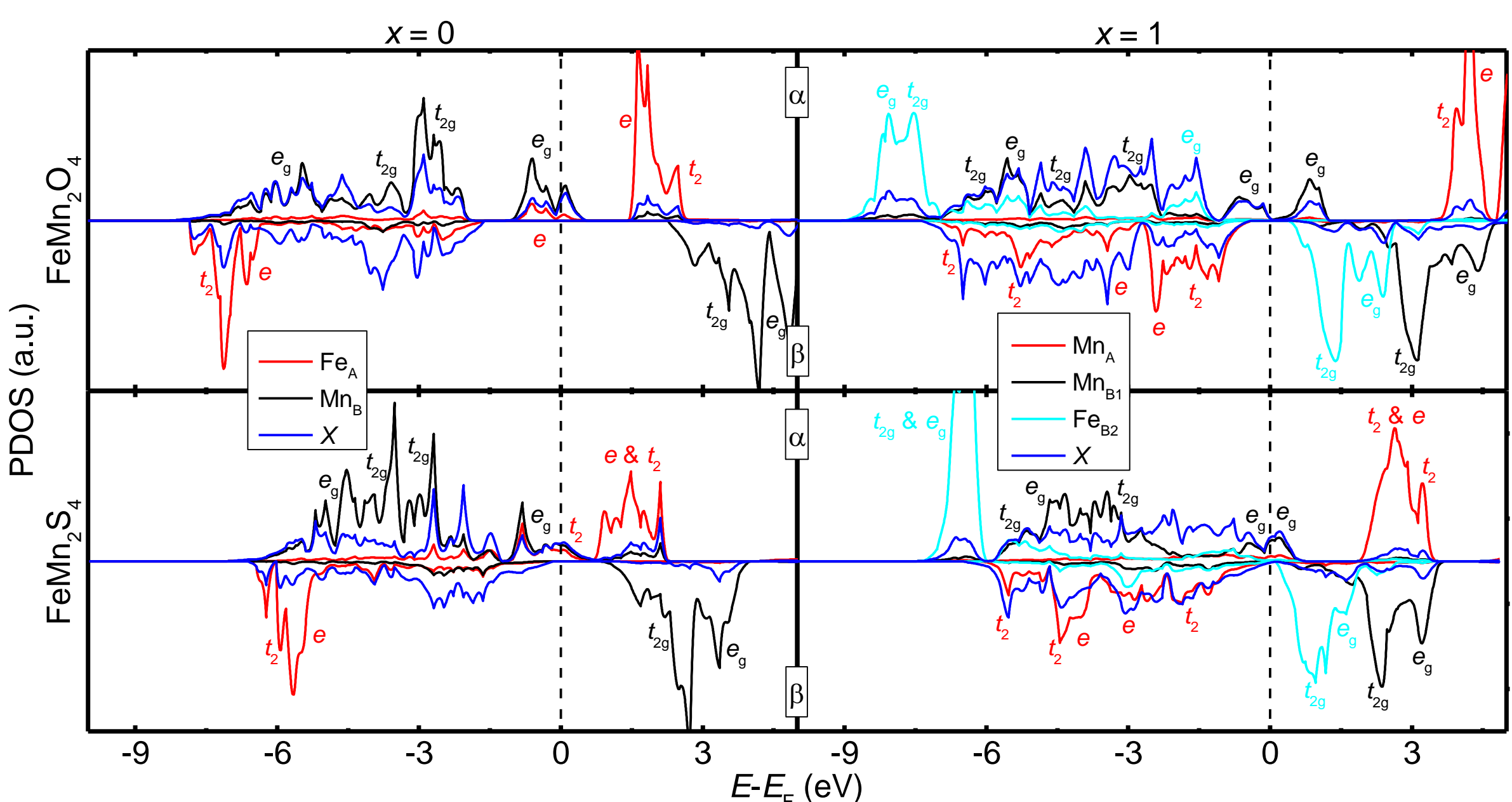

FIG. 4. Atomic projections of the spin decomposed total density of states (PDOS) for $FeMn_2O_4$ and $FeMn_2S_4$. Fe and Mn contributions are from the 3$d$ bands. O and S contributions are from the 2$p$ and 3$p$ orbitals respectively.

### *3. $FeCo_2X_4$*

When $FeCo_2O_4$ has a normal distribution, all the bands are pushed slightly towards the Fermi energy and especially those due to $Co_B$, see Fig. 5 top-left panel. As a result, the partially-occupied $Co_B$ $t_{2g}$ level that crosses the Fermi energy has a minimal band gap in the minority channel of the spins, making the normal $FeCo_2O_4$ spinel almost a half-metal, see also the integer value of the spin magnetization of saturation in Table IV. On the other hand, the sulfide counterpart has all the bands closer to the Fermi energy with symmetrical $Co_B$ bands in the minority and majority spin channels (Fig. 5 bottom-left panel), due to the fully-occupied $t_{2g}$ level which indicates non-magnetic behaviour for this atom (see Table IV). There is a peak at the Fermi energy, in the majority spin channel, with contributions from the partially-occupied $Fe_A$ $e$ and $Co_B$ $e_g$ levels. In the minority spin channel, the normal $FeCo_2S_4$ spinel is weakly conducting, as there is a small $Co_B$ fully-occupied $t_{2g}$ band strongly hybridized with S $p$ orbitals that ends shortly after the Fermi energy in the conduction band side. Overall, the sulfide counterpart is metallic which is confirmed by a decimal spin magnetization of saturation, Table IV.

When the Co-based (thio)spinels have an inverse cation distribution, all the bands are slightly pushed away from the Fermi energy compared to the normal cation distribution, especially in the oxide, see Fig. 5 right panels. $Fe_{B2}$ $d$ bands appear in the typical range described so far for both oxide and sulfide spinels. For the oxide, $Co_A$ valence $d$ bands are in both spin channels, while in the α spin channel the partially-occupied $t_2$ level appears exclusively in the conduction part. The fully-occupied $Co_{B1}$ valence $t_{2g}$ levels are nearly symmetrically placed in both spin channels, rendering this atom as non-magnetic, see $m_s$ in Table IV. The inverse cation distribution of the oxide has insulating properties, see also $M_S$ in Table IV. Although the sulfide counterpart has the $Co_A$ and $Co_{B1}$ bands symmetrically placed in both channels of the spins, the bands crossing the Fermi energy give it metallic properties (see the decimal value of the spin magnetization of saturation $M_S$ = 2.46 $\mu_B$/f.u. in Table IV, typical of metals). These properties are due to the hybridized partially-occupied $Co_A$ $t_2$ and fully-occupied $Co_{B1}$ $t_{2g}$ levels and to the merged $Co_A$ $t_2$, $Co_{B1}$ $t_{2g}$ and $Fe_{B2}$ $t_{2g}$ levels in the majority and minority channel of spins, respectively.

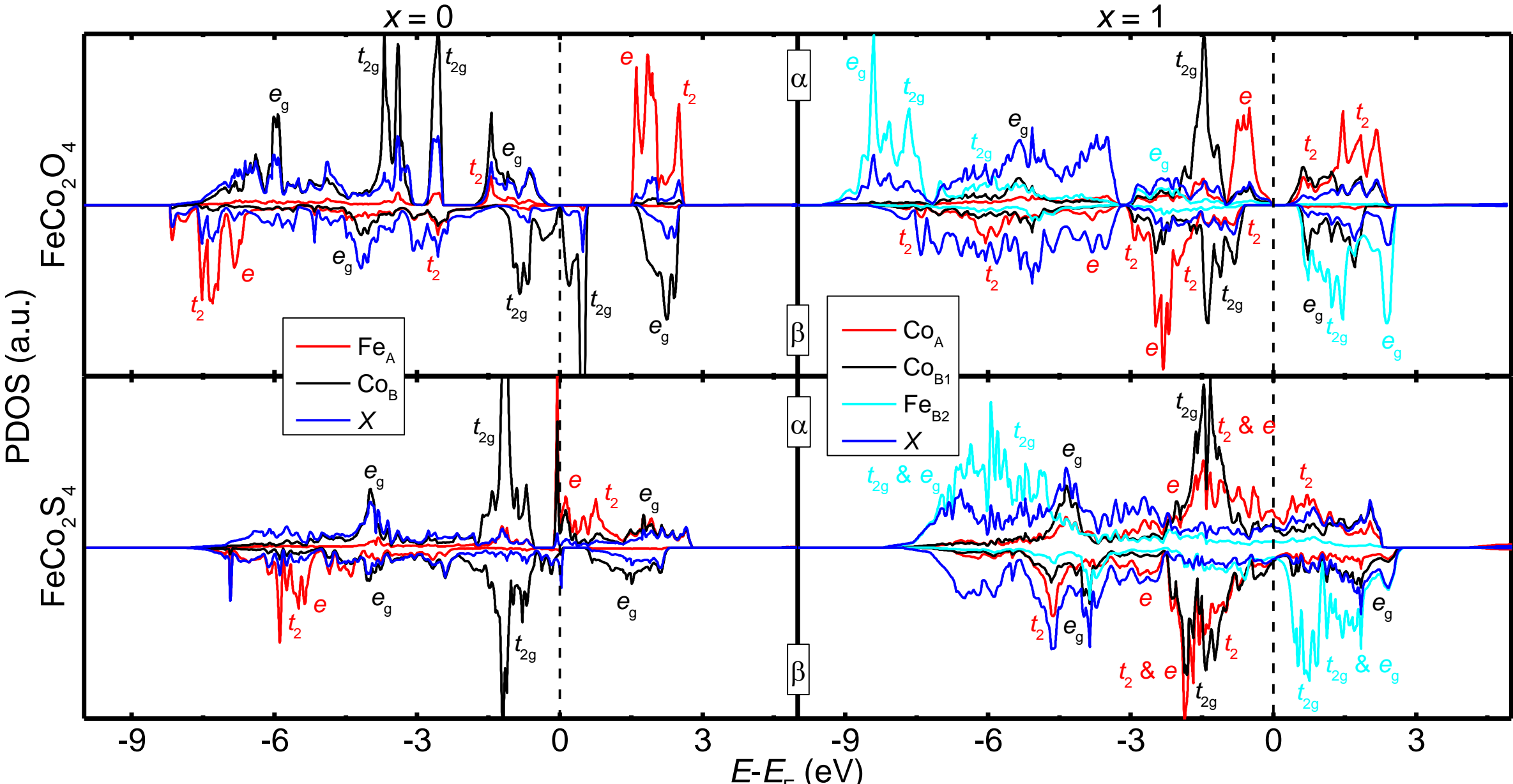

FIG. 5. Atomic projections of the spin decomposed total density of states (PDOS) for $FeCo_2O_4$ and $FeCo_2S_4$. Fe and Co contributions are from the 3*d* bands. O and S contributions are from the 2*p* and 3*p* orbitals respectively.

### *4. $FeNi_2X_4$*

For both Ni-based (thio)spinels, when $x = 0$, the bands' pattern is similar and follows the same distribution as described in previous cases, see Fig. 6 left panels. The oxide is half-metal due to a strong hybridization of the partially-occupied $Ni_B$ $e_g$ band with the O *p* orbitals that cross the Fermi energy in the majority channel of the spins, see also the integer value of $M_S$ in Table IV. The main difference between oxide and sulfide lies in the fact that bands in the latter are closer to the Fermi energy in both spin channels, thus becoming a metallic system. The thiospinel's metallic character is given by the S *p* orbitals with a small hybridization (only in the majority channel of the spins) with the partially-occupied $Fe_A$ *e* and $t_{2g}$ levels, see the decimal $M_S$ in Table IV.

When Ni is filling the A sublattice, the Ni-based spinel becomes half-semiconductor with a band gap of 0.20 and 2.05 eV in the majority and minority channels of the spins, respectively, see Fig. 6 top right panel. While the position and distribution of the bands due to the ions occupying different positions is equivalent to what we have presented in previous cases, for inverse $FeNi_2O_4$ spinel, nonetheless, this is not the case for the sulfide counterpart. In the inverse $FeNi_2S_4$ system, the $Ni_A$ and $Ni_{B1}$ ions are less magnetic than expected and the valence and conduction bands are merged together, making this compound metallic for any inversion degree. The inverse Ni-based (thio)spinels have integer values of $M_S$ (see Table IV) regardless of whether they are insulator or metal. Note that in the case of the metal inverse $FeNi_2S_4$ spinel, the decimal number in $M_S$ = 4.00 $\mu_B$/f.u. is a special case. In $FeNi_2S_4$, for any inversion degree, the metallic character agrees with the experimental findings.[21,103]

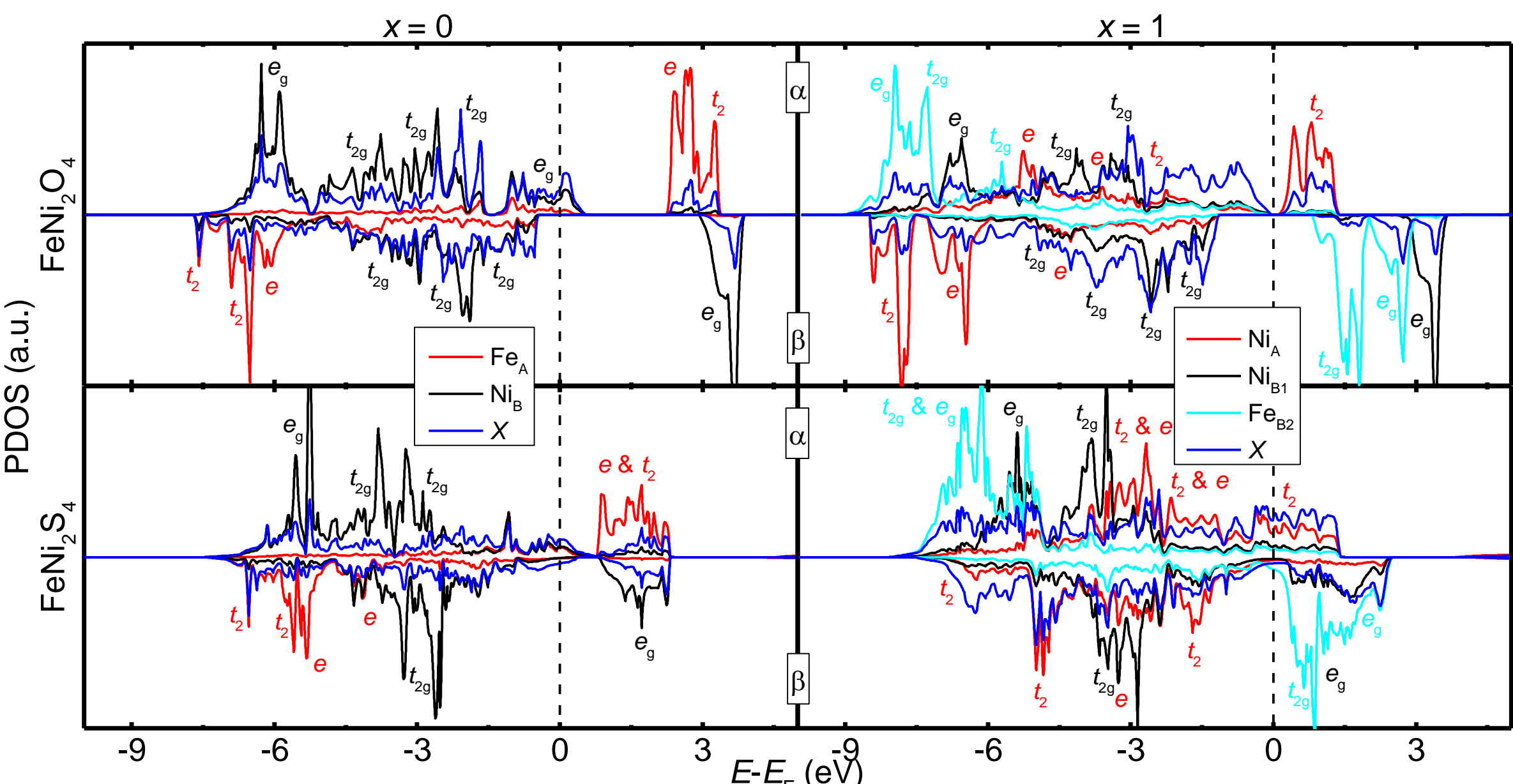

FIG. 6. Atomic projections of the spin decomposed total density of states (PDOS) for $FeNi_2O_4$ and $FeNi_2S_4$. Fe and Ni contributions are from the 3*d* bands. O and S contributions are from the 2*p* and 3 *p* orbitals respectively.

## IV. SUMMARY AND CONCLUSIONS

We have performed systematic electronic structure calculations for a series of (thio)spinels, which elucidate the cation distribution as well as the magnetic and electronic properties of these materials.

We have determined the thermodynamic inversion degree for the $FeM_2X_4$ (thio)spinels at temperatures used typically in their synthesis, which agrees reasonably well with available experimental evidence. More quantitative results could be expected if additional values of inversion degrees and different cations arrangements to those explored in this work are considered for the spinel compositions, although we do not expect this to change the trend of our results. We have found that $FeM_2X_4$ spinels are more likely to have a normal distribution of cations when $M$ is one of the two atoms to the left of Fe in the periodic table. $FeMn_2O_4$ has a metastable intermediate inversion degree that could only be found by considering entropic factors, which also agrees with experiment. It may be that the global minimum, *i.e.* the inverse spinel, is difficult to attain due to kinetic control. When $M$ is one of the two atoms to the right of Fe in the periodic table, with the exception of $FeCo_2O_4$, the spinels have an intermediate inversion degree ranging between 10 to 50%. Finally, the oxidic spinel of Co and Fe has a completely inverse distribution of the cations. The small equilibrium inversion degree of $FeNi_2S_4$ agrees acceptably well with the one found in natural samples. Fitting the experimental spin magnetization of saturation of $FeNi_2X_4$ to the ones calculated for the normal and inverse structures gives inversion degrees with a similar trend to those calculated using thermodynamic arguments. This procedure could also be applied to $FeCo_2S_4$ and $FeNi_2O_4$ if the magnetizations of saturation are known experimentally.

No single factor among those analyzed, *i.e.* neither crystal field stabilization effects nor the size of the cations, can account by themselves for the equilibrium inversion degree.

For the two extreme scenarios, namely the completely normal and inverse spinels, we have calculated the electronic and magnetic properties of the metal atoms as well as the electronic properties of the bulk phase. We found that the majority of the spinels for any extreme inversion degree are half-metals in the ferrimagnetic state. Notable exceptions are the inverse Co and Ni oxide spinels, which are insulators, and their sulfide counterparts that are metallic for any inversion degree, together with the inverse $FeMn_2S_4$. Notably, we found that hard anions stretch the band structure, giving the biggest band gaps and therefore the best half-metallic properties.

Finally, we have proposed a theoretical structure for $FeMn_2S_4$ and $FeCo_2S_4$ and have predicted their electronic and magnetic properties and equilibrium inversion degree.

## ACKNOWLEDGEMENTS

We acknowledge the Engineering & Physical Sciences Research Council (grants EP/G036675 and EP/K035355) for funding. *Via* our membership of the UK's HPC Materials Chemistry Consortium, which is funded by EPSRC (EP/L000202), this work made use of the facilities of HECToR and ARCHER, the UK's national high-performance computing services, which are funded by the Office of Science and Technology through EPSRC's High End Computing Programme. The authors also acknowledge the use of the UCL *Legion* High Performance Computing Facility (Legion@UCL), and associated support services, in the completion of this work. The authors would like to acknowledge that the work presented here made use of the IRIDIS High Performance Computing facility provided *via* the Centre for Innovation (CfI), comprising the universities of Bristol, Oxford, Southampton and UCL in partnership with the STFC Rutherford Appleton Laboratory. DS-C thanks UCL for a Graduate Global Excellence Award and an Overseas Research Scholarship from the UCL Industrial Doctorate Centre in Molecular Modelling and Materials Science. AR is grateful to the Ramsay Memorial Trust and University College London for the provision of a Ramsay fellowship, and NHdL acknowledges the Royal Society for an Industry Fellowship.